\documentclass{jaa}
\usepackage{threeparttable}
\usepackage{amsmath}
\usepackage[utf8]{inputenc}
\usepackage{longtable}
\usepackage{subcaption}
\usepackage{lipsum}
\usepackage{graphicx}
\usepackage{natbib}
\usepackage{multirow}
\usepackage[Table,xcdraw]{xcolor}
\usepackage{rotating}
\newcommand{\apj}{Astrophysical Journal}

\newcommand{\mnras}{Monthly Notices of the Royal Astronomical Society}

\newcommand{\aap}{Astronomy and Astrophysics}

\usepackage{hyperref}
\hypersetup{backref=true,
    pagebackref=true,
    hyperindex=true,
    colorlinks=true,
    breaklinks=true,
    urlcolor= black,
    linkcolor= blue,
    bookmarks=true,
    bookmarksopen=false,
    filecolor=black,
    citecolor=blue,
    linkbordercolor=blue
}

\begin{document}\sloppy

\title{Analyses of Hydrogen--stripped core--collapse supernovae using MOSFiT and MESA based tools}


\author{Amar Aryan\textsuperscript{1,2}, Shashi Bhushan Pandey\textsuperscript{1}, Amit Kumar\textsuperscript{1,3}, Rahul Gupta\textsuperscript{1,2}, Amit kumar Ror\textsuperscript{1}, Apara Tripathi\textsuperscript{2}, and Sugriva Nath Tiwari\textsuperscript{2}}

\affilOne{\textsuperscript{1}Aryabhatta research institute of observational sciences (ARIES), Nainital, Uttarakhand, India, \\}

\affilTwo{\textsuperscript{2}Department of Physics, Deen Dayal Upadhyaya Gorakhpur University, Gorakhpur-273009, India \\}


\affilThree{\textsuperscript{3}School of Studies in Physics and Astrophysics, Pt. Ravishankar Shukla University, Chattisgarh 492010, India\\}


\twocolumn[{

\maketitle

\corres{amararyan941@gmail.com, amar@aries.res.in}

\msinfo{---}{---}

\begin{abstract}
{In this work, we employ two publicly available analysis tools to study four hydrogen(H)--stripped core--collapse supernovae (CCSNe) namely, SN~2009jf, iPTF13bvn, SN~2015ap, and SN~2016bau. We use the Modular Open-Source Fitter for Transients ({\tt MOSFiT}) to model the multi band light curves. {\tt MOSFiT} analyses show ejecta masses (log\,M$_{ej}$) of $0.80_{-0.13}^{+0.18}$\,M$_{\odot}$, $0.15_{-0.09}^{+0.13}$\,M$_{\odot}$, $0.19_{-0.03}^{+0.03}$\,M$_{\odot}$, and $0.19_{+0.02}^{-0.01}$\,M$_{\odot}$ for SN~2009jf, iPTF13vn, SN~2015ap, and SN~2016au, respectively. Later, Modules for Experiments in Stellar Astrophysics ({\tt MESA}), is used to construct models of stars from pre-main sequence upto core collapse which serve as the possible progenitors of these H-stripped CCSNe. Based on literature, we model a 12\,M$_{\odot}$ ZAMS star as the possible progenitor for iPTF13vn, SN~2015ap, and SN~2016bau while a 20\,M$_{\odot}$ ZAMS star is modeled as the possible progenitor for SN~2009jf. Glimpses of stellar engineering and the physical properties of models at various stages of their lifetime have been presented to demonstrate the usefulness of these analysis threads to understand the observed properties of several classes of transients in detail.}

\end{abstract}

\keywords{Supernovae---{MOSFiT}---{\tt MESA}}

}]


\doinum{}
\artcitid{\#\#\#\#}
\volnum{000}
\year{0000}
\pgrange{1--}
\setcounter{page}{1}
\lp{11}

\section{Introduction}
\label{Intro}

Core-collapse Supernovae (CCSNe) are extremely powerful explosions that mark the death of massive stars, further sub-divided into various classes subjected to the presence/absence 
of H in their photospheric phase spectra \citep[][]{Filippenko1988, Filippenko1993, Smartt2009, Maoz2014,VanRossum2016,Konyves2020}. Among H-deficient ones, near peak the Type Ib SNe exhibit prominent Helium (He)--features in their spectra, whereas Type Ic SNe show neither H nor He obvious features. Prominent features of intermediate-mass elements such as O, Mg, and Ca are also seen in Type Ib and Type Ic SNe spectra. Another very interesting class, known as Type IIb SNe form a transition class of objects that are supposed to link SNe~II and SNe~Ib \citep[][]{Filippenko1988, Filippenko1993, Smartt2009}. The early-phase spectra of SNe~IIb display prominent H--features, while unambiguous He--features appear after a few weeks \citep[][]{Filippenko1997}. The CCSNe are the final destinations of massive stars ($\gtrsim$ 8--10\,$M_\odot$;  e.g., \citealt[][]{Garry2004, Woosley2005, Groh2017}), resulting from the core-collapse due to the exhaustion of the nuclear fuel in their cores.

Underlying physical mechanisms behind above-mentioned classes of CCSNe are still not understood well. One popular mechanism is the neutrino--driven outflow \citep[][and references therein]{Muller2017} but other mechanisms have also been proposed (e.g., magnetorotational mechanism of the explosion of CCSNe as discussed in \citet[][]{Bisnovatyi2018}). In many cases, the CCSNe explosions are not spherically symmetrical. As a particular case, studies by \citet[][]{Couch2009} indicate that the aspherical CCSNe from red supergiants are powered by non-relativistic Jets. Further, \citet[][and references therein]{Piran2019} mention that gamma-ray bursts (GRBs) that accompany rare and powerful CCSNe (popularly known as 'hypernovae') involve the association of relativistic jets emerging due to the explosion of a certain class of massive stars exploding under specific physical conditions.

\begin{figure*}
\includegraphics[scale=0.31]{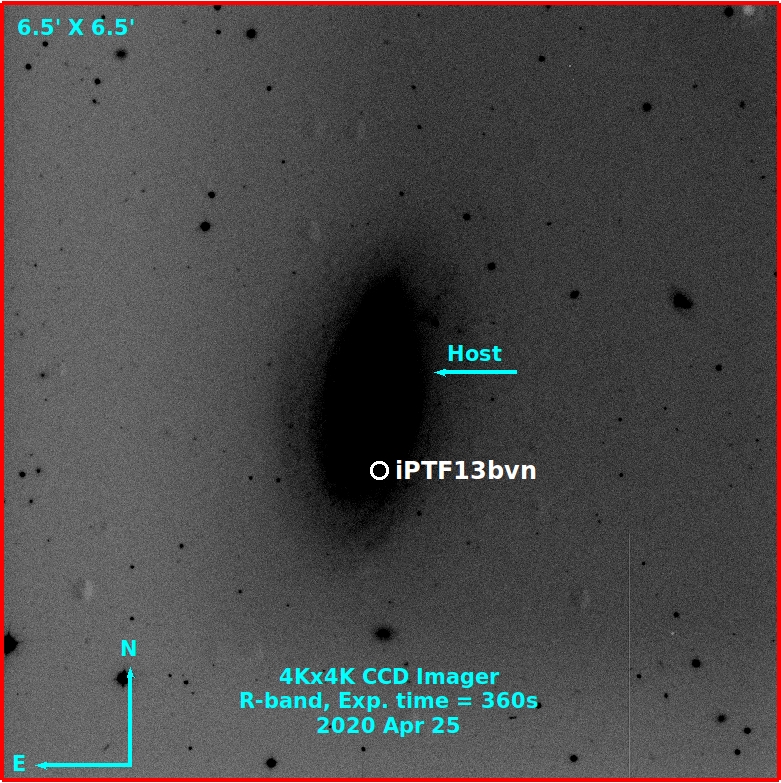}
\includegraphics[scale=0.35]{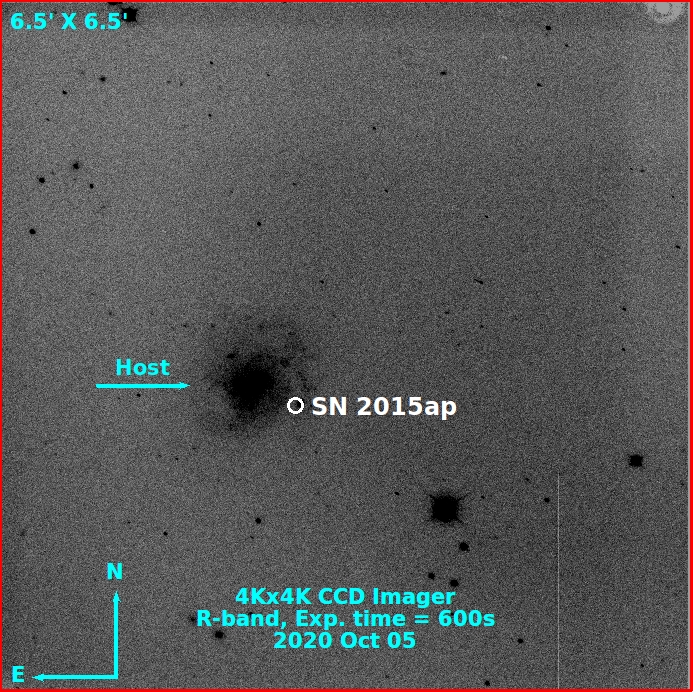}
\caption{Finding charts of iPTF13bvn and SN 2015ap host galaxies in the $R$-band using the 4K$\times$4K CCD IMAGER \citep{Pandey2018, Kumar2021} mounted at the axial port of the 3.6m DOT in the left, and right panels respectively, are shown to demonstrate a diverse set of environments and locations within the host galaxies.}
\label{fig:finding_charts}
\end{figure*}

During the last stages of their lives, the stars are fully evolved and contain mostly intermediate (e.g. Si, Mg etc.) to high mass elements (e.g. Fe, Ni, Co etc.) through various nuclear processes. So, stellar deaths through these catastrophic events are also responsible for the chemical enrichment of our universe apart from the birth of compact objects like neutron stars and black holes only. Nuclear--astrophysics aims at understanding the nuclear processes that take place in the universe. These nuclear processes generate energy in stars and contribute to the nucleosynthesis of the elements and the evolution of the galaxy through many possible channels including SN explosions. The collapses of the stellar cores produce elements through various possible ways (e.g. s/r/p processes) enriching interstellar medium. This research field has now evolved as a developed one  \citep[e.g.][and references therein]{Liccardo2018,Meyer2020} demanding a multi-wavelength study to better understand not only the nuclear processes but also the nature of the possible progenitors behind such cosmic explosions. It is now known that a good fraction of well-studied H-stripped CCSNe happen in diverse types of host galaxies (see figure~\ref{fig:finding_charts}) having a range of physical properties like mass, luminosity, age, star formation rates, etc. Late time observations of these host galaxies are of crucial importance to decipher the nature of possible progenitors exploding in a diverse range of environments. Apart from multi-band observations of these events, detailed studies about ambient media of such host galaxies and their pre-explosion images are very useful to constrain the nature of possible progenitors.

\begin{table*}
\caption {The adopted total-extinction values (E(B-V)$_{tot}$), adopted luminosity distances (D$_L$), and redshift (z) of SNe considered in the present analysis.}
\label{tab:comparison_Sample}
\begin{center}
{\scriptsize
\begin{tabular}{ccccccccccccc}
\hline\hline
	SN name    &	$E(B-V)_{\rm tot}$	&	(D$_L$)  & redshift \\
               &	(mag)	&	(Mpc) &  &  \\
\hline
SN~2009jf  	 	&	0.112 \citep[][]{Sahu2011}	 & 34.66\,Mpc \citep[][]{Sahu2011} &  0.007942 \citep[][]{Sahu2011}\\

iPTF13bvn 	 	&	0.21 \citep[][]{Bersten2014} & 25.8\,Mpc \citep[][]{Bersten2014} &  0.00449 \citep[][]{Cao2013}\\

SN~2015ap 	 	&	0.037 \citep[][]{Prentice2019,Aryan2021a}   & 46.6\,Mpc \citep[][]{Aryan2021a} &  0.01138 \citep[][]{Aryan2021a}\\

SN~2016bau  	    &	0.579 \citep[][]{Aryan2021a}   & 21.77\,Mpc \citep[][]{Aryan2021a} &  0.003856 \citep[][]{Aryan2021a}\\
\hline\hline
\end{tabular}}
\end{center}
\end{table*}

In the light of above and the published work earlier, as part of the present analysis we chose a sample of 4 H-stripped CCSNe (i.e. SN~2009jf \citep[][]{Sahu2011,Valenti2011}, iPTF13bvn \citep{Cao2013, Bersten2014,Eldridge2015}, SN~2015ap \citep[][]{Aryan2021a}, and SN~2016bau \citep[][]{Aryan2021a}) to fit their multi-band optical light curves using {\tt MOSFiT} \citep[][]{Guillochon2018}. Further, the fitting parameters obtained using {\tt MOSFiT} are used to characterize the nature of possible progenitors using 1-dimensional stellar evolution code {\tt MESA} \citep[][] {Paxton2011,Paxton2013,Paxton2015,Paxton2018,Paxton2019}. All the analyses performed in this work make use of publicly available tools. More details about the usefulness of analysis tools like {\tt MESA/MOSFiT} (and others) are described in \citet[][among many other]{Aryan2021b,Aryan2022} depicting how these tools can be boon to the transient community.    

This paper has been divided into five sections. A brief introduction and methods to fit the light curves of various SNe have been presented in section~\ref{mosfit}. In section~\ref{Mesa}, the basic assumptions and various physical and chemical properties of the possible progenitors of different SNe have been presented. We discuss the major outcomes of our studies in section~\ref{Results} and provide our concluding remarks in section~\ref{Conclusion}. 

\begin{figure*}
  \centering
  \begin{tabular}{cc}
    \subcaptionbox{\label{Picture}}[\columnwidth]{\includegraphics[width=\columnwidth]{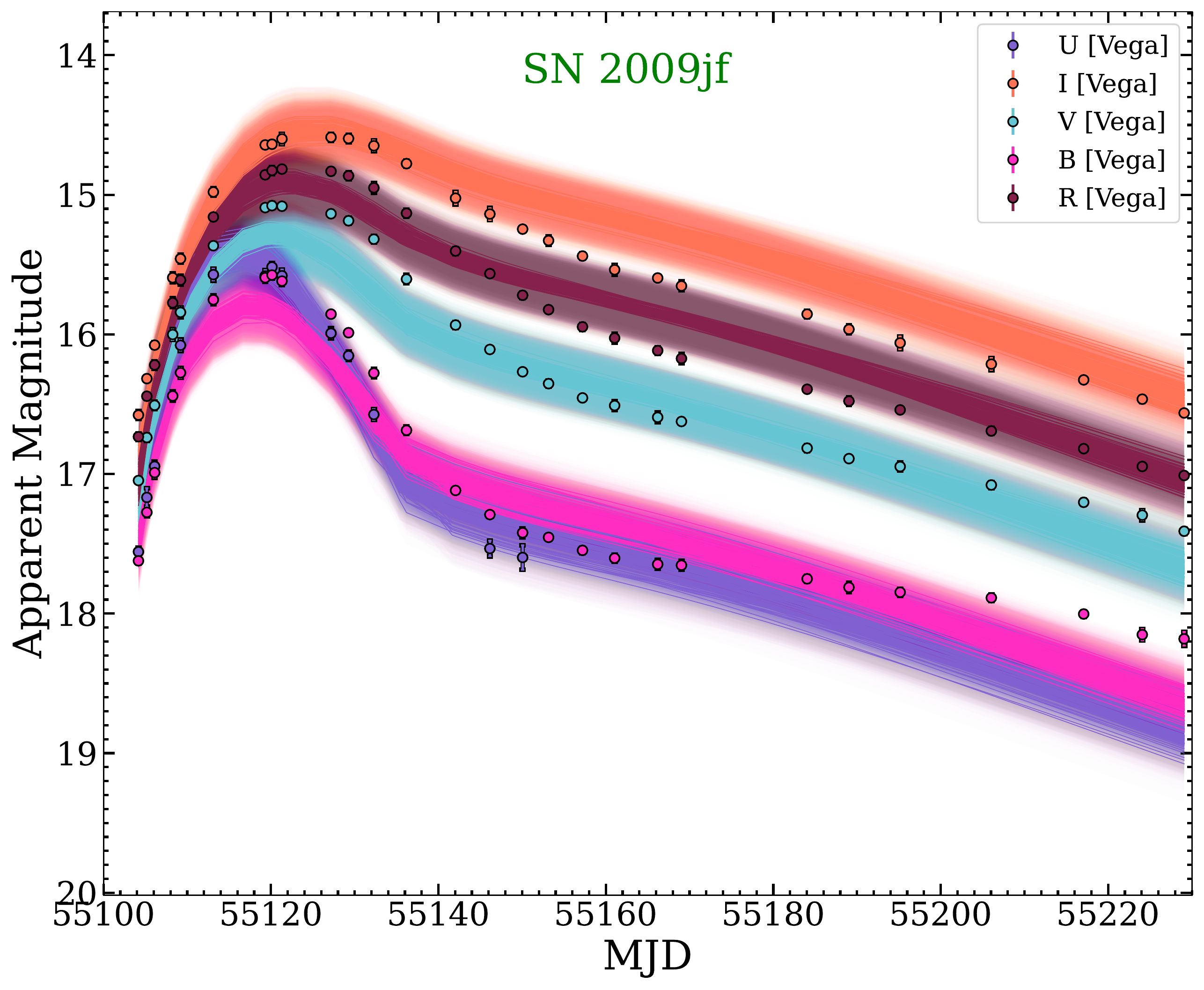}}
    \subcaptionbox{\label{PictureA}}[\columnwidth]{\includegraphics[width=\columnwidth]{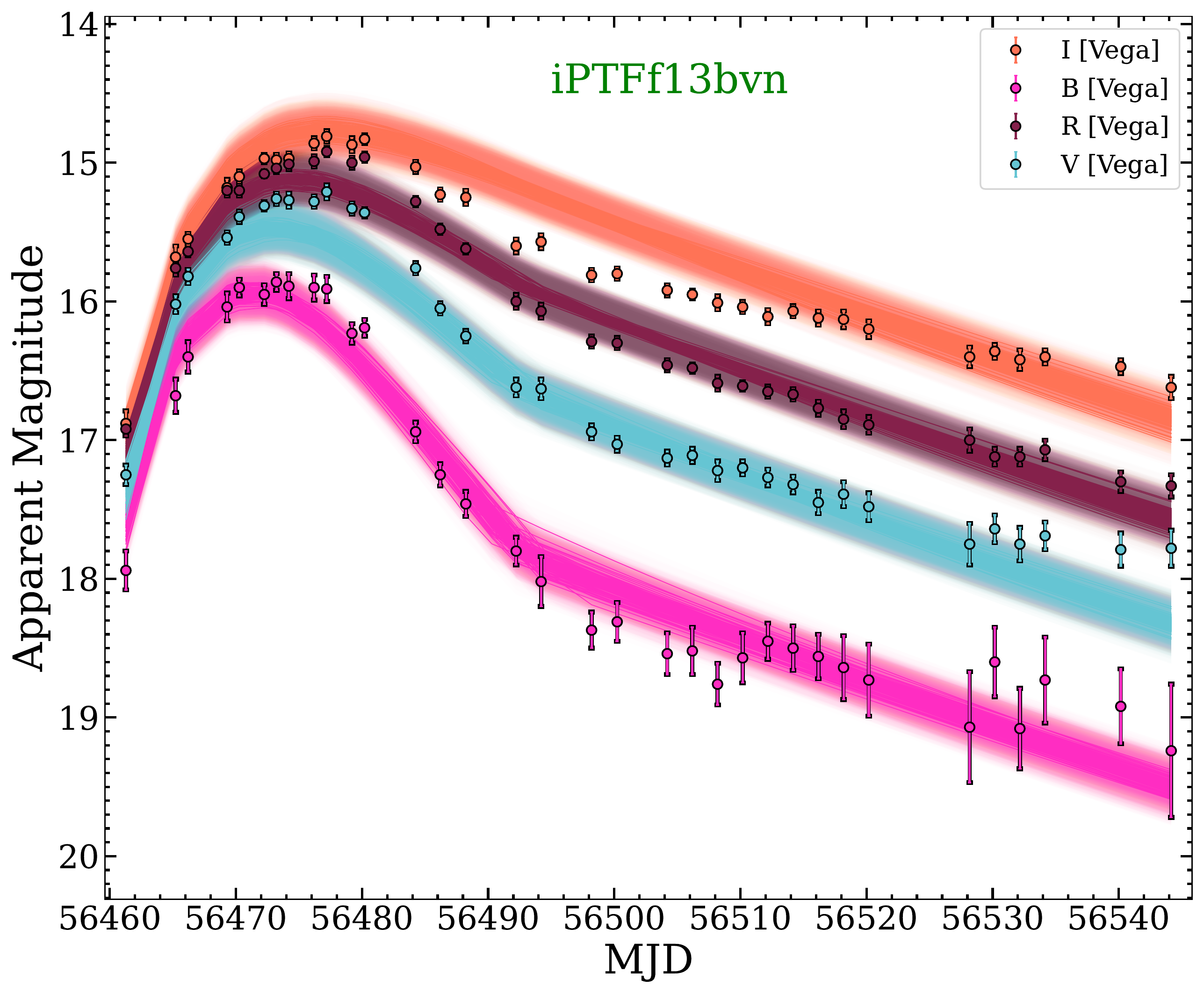}}\\
    \subcaptionbox{\label{PictureB}}[\columnwidth]{\includegraphics[width=\columnwidth]{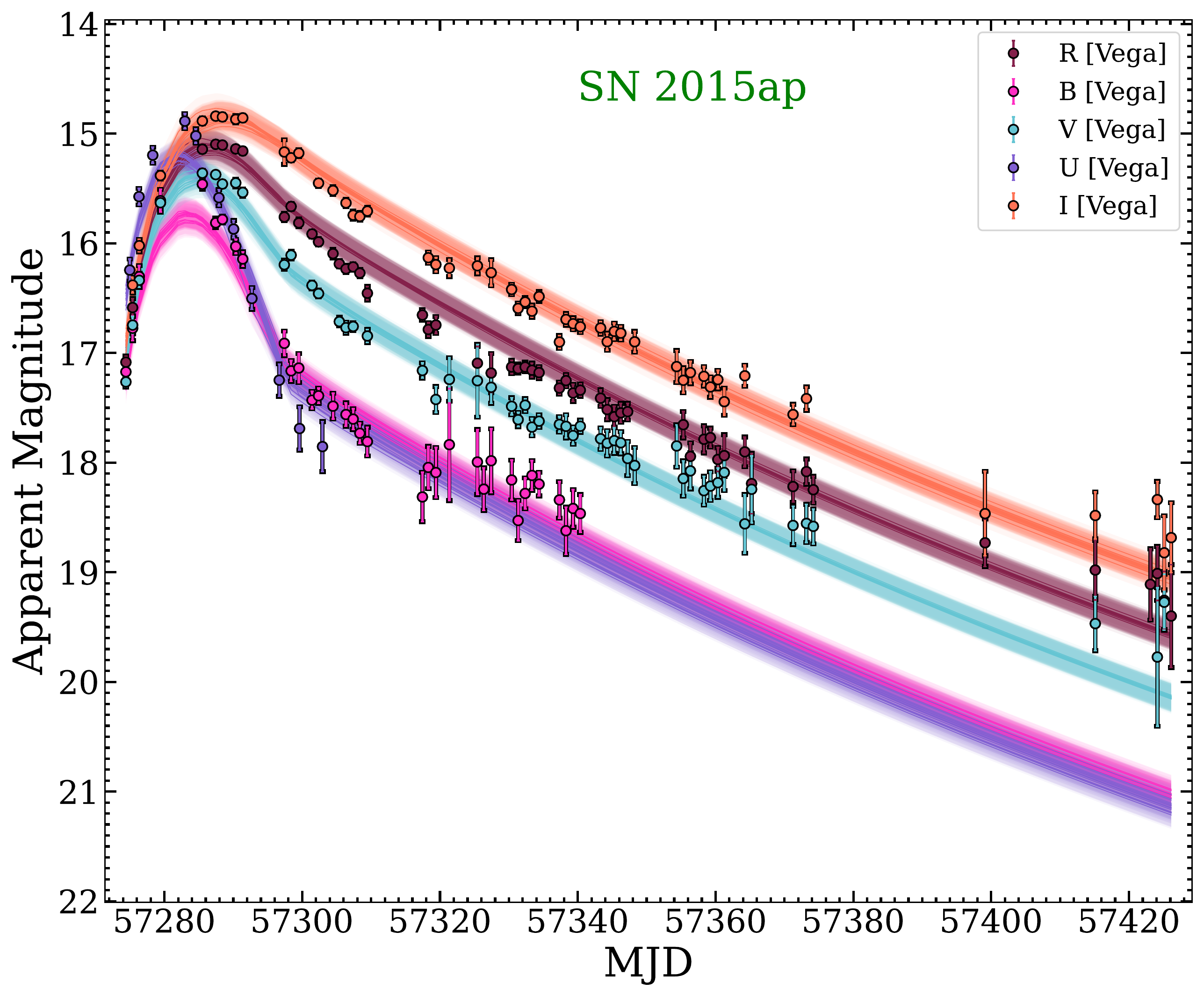}}
    \subcaptionbox{\label{PictureC}}[\columnwidth]{\includegraphics[width=\columnwidth]{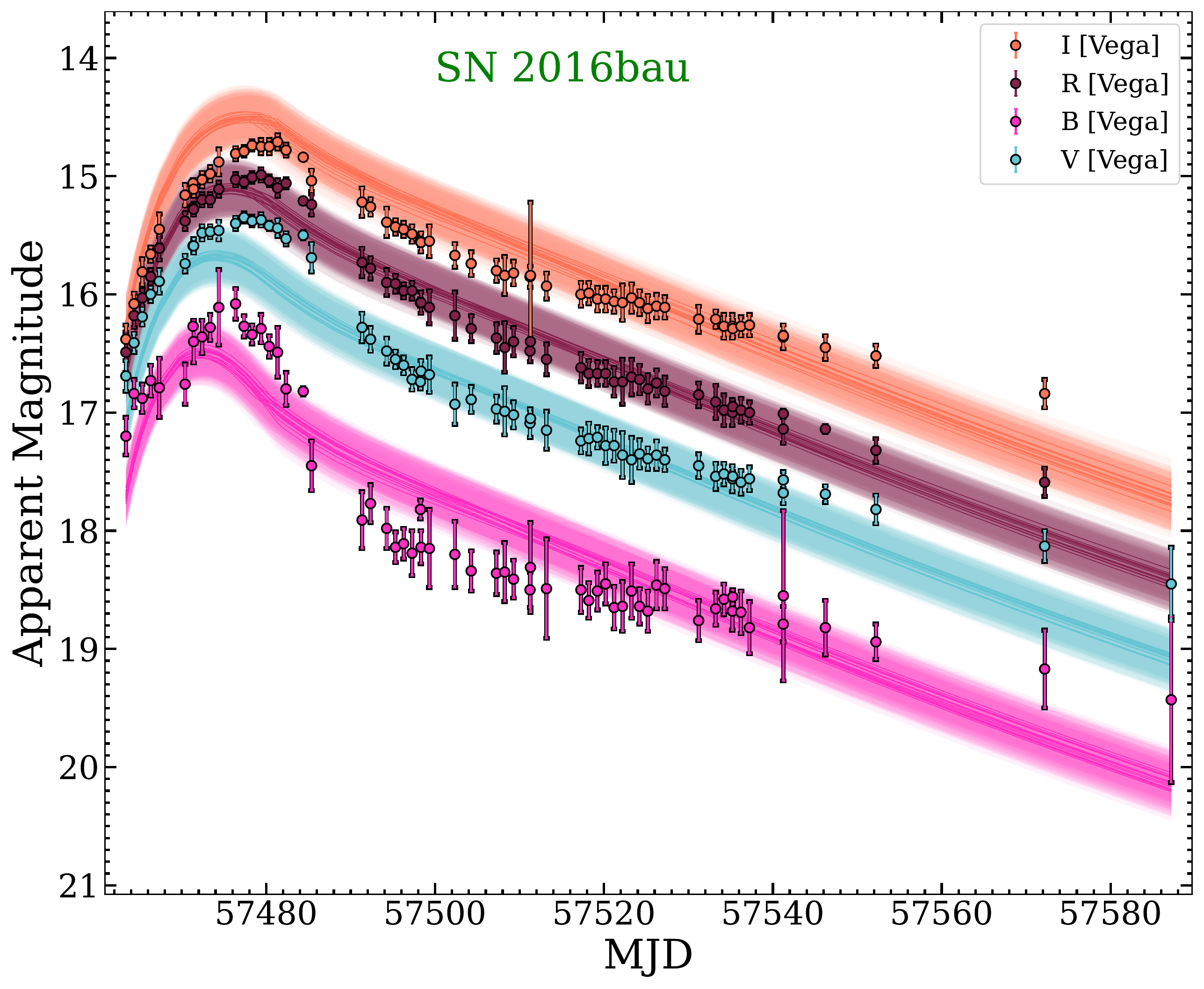}}
  \end{tabular}
  \caption{The results of {\tt MOSFiT} fittings to the multi band light curves of SN~2009jf, iPTF13bvn, SN~2015ap, and SN~2016bau, respectively. For type Ib SNe, the radioactive decay of $^{56}$Ni and $^{56}$Co is considered to be the prominent powering mechanism for their light curves. Thus, the {\tt default} model from {\tt MOSFiT} has been employed to fit the light curves of these SNe.}
  \label{fig:mosfit_typeIb}
\end{figure*}

\section{Fitting the light curves using {\tt MOSFiT}} 
\label{mosfit}
{\tt MOSFiT} is a Python-based package that downloads data from openly available online catalogs, generates the Monte Carlo ensembles of semi--analytical light-curve fits to downloaded data sets along with their associated Bayesian parameter posteriors and provides the fitting results back. Besides fitting data downloaded from openly available online catalogs, one can also perform a similar analysis to the private data sets. {\tt MOSFiT} employs various powering mechanisms for the light curves of different types of SNe. Some of them are; a) {\tt default} model incorporating the Nickel-Cobalt decay \citep[][]{Nadyozhin1994}, b) {\tt magnetar} model that takes a magnetar engine with simple spectral energy distribution \citep[][]{Nicholl2017}, and c) {\tt csm} model which are interacting CSM--SNe \citep[][]{Chatzopoulos2013, Villar2017}. A detailed description of all the models available through {\tt MOSFiT} is provided in \citet[][]{Guillochon2018}.

In this work, we tried to fit the multi-band light curves of four Type Ib SNe. These four SNe are SN~2009jf, iPTF13bvn,  SN~2009jf, and SN~2016bau. We employ the {\tt default} model to fit the multi-band light curves of these SNe. The multi-band light curves of SN~2009jf and iPTF13bvn have been taken from \citet[][]{Sahu2011} and \citet[][]{Bersten2014}, respectively, while the source of multi-band light curves of SN~2015ap and  SN~2016bau is \citet[][]{Aryan2021a}. Further, total extinction (E(B-V)$_{tot}$), luminosity distance (D$_L$) and redshift (z) of each SN in the present study are given in Table~\ref{tab:comparison_Sample}. 

The {\tt MOSFiT} calculations for all other four SNe were already performed in \citet[][]{Meyer2020}, but in their study, \citet[][]{Meyer2020} simply fit the data available from Open supernova catalog \citep[][]{Guillochon2017}. Previous studies and fits performed from the data directly imported from open supernova catalog lack a few corrections including host galaxy extinction correction etc. Also, sometimes only a few data points are available to fit from open supernova catalog similar to the cases like SN~2015ap and SN~2016bau in \citet[][]{Meyer2020}. However, in the present analysis, the data for each SN are taken from the referenced sources, corrected for total (Milky way and host galaxy) extinction, and then privately fit using {\tt MOSFiT} using a larger data set. Thus, we consider our fittings using {\tt MOSFiT} to be more reliable, and thus fitting parameters obtained through our analyses are better.     

For SN2009jf, {\tt MOSFiT} fittings give an ejecta mass of (M$_{ej}$) $\sim$ 6.31\,M$_{\odot}$, which is in very good agreement with \citet[][]{Sahu2011} and \citet[][]{Valenti2011}, while \citet[][]{Meyer2020} under-predicts the M$_{ej}$ ($\sim$ 2.45\,M$_{\odot}$). Further, for iPTF13bvn, we obtain M$_{ej}$ $\sim$ 1.41\,M$_{\odot}$ agreeing closely with \citet[][]{Eldridge2015} and \citet[][]{Paxton2019}, while \citet[][]{Meyer2020}, once again under-predicts the M$_{ej}$. Similarly, for SN~2015ap and SN~2016bau, we obtain ejecta masses of $\sim$ 1.54\,M$_{\odot}$ and 1.54\,M$_{\odot}$, respectively. The M$_{ej}$ from our {\tt MOSFiT} fittings for SN~2015ap is close to \citet[][]{Prentice2019} but lower than what is obtained from \citep[][]{Aryan2021a}. For SN~2015ap, \citet[][]{Meyer2020} also produced similar value of M$_{ej}$. The M$_{ej}$ from our {\tt MOSFiT} fittings for SN~2016bau are very close to \citep[][]{Aryan2021a} while \citet[][]{Meyer2020} predicts much lower value.

\begin{figure*}[!t]
  \includegraphics[height=20cm,width=18cm]{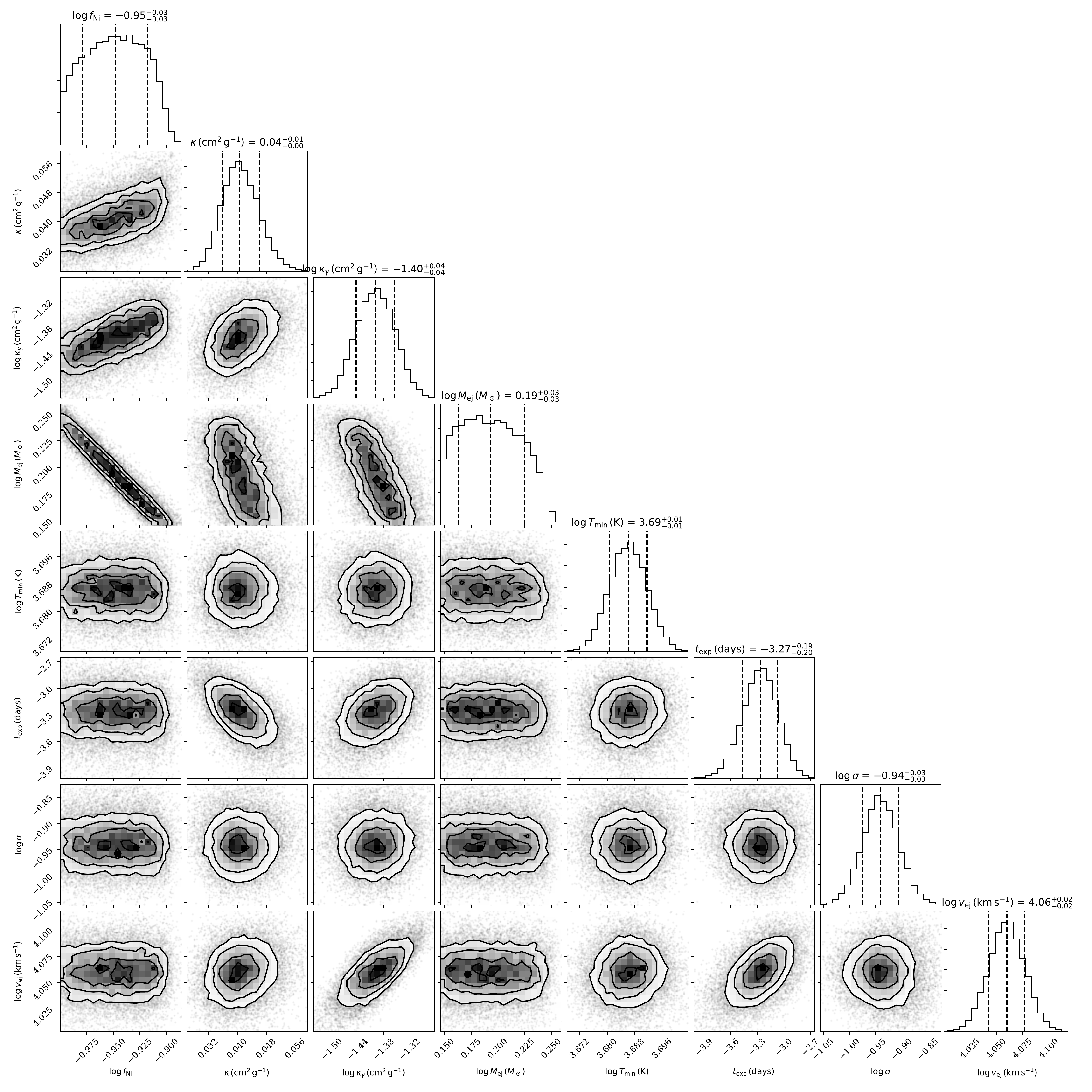}
  \caption{The corner plot of the fitting parameters from the {\tt default} model for SN~2015ap using {\tt MOSFiT}.}
  \label{fig:corner_SN2015ap}
\end{figure*}

Figure~\ref{fig:mosfit_typeIb} shows the results of fitting {\tt default} model using {\tt MOSFiT} to the multi band light curves of SN~2009jf, iPTF13bvn, SN~2015ap, and SN~2016bau. The corner plot of the fitting parameters of the {\tt default} model for SN~2015ap has been shown in figure~\ref{fig:corner_SN2015ap} as an example. Similar corner plots are generated for other SNe also.  The fitting parameters of the {\tt default} model to other four SNe are listed in Table~\ref{tab:mosfit_default}. 


\begin{table*}
\centering
\begin{scriptsize}
\begin{tabular}{l|ccccccccc}
\hline
\hline
Source name & $\log\, M_{\rm ej}\,(M_\odot)$ & $\log\, f_{\rm Ni}$ & $\kappa\,({\rm cm}^{2}\,{\rm g}^{-1})$ & $\log\, \kappa_\gamma\,({\rm cm}^{2}\,{\rm g}^{-1})$ & $\log\, v_{\rm ej}\,({\rm km\,s}^{-1})$ & $\log\, T_{\min}\,{\rm (K)}$ & $\log\, \sigma$ & $t_{\rm exp}\,{\rm (days)}$ \\
\hline
\\
SN2009jf & $0.80_{-0.13}^{+0.18}$ & $-1.47_{+0.13}^{-0.18}$ & $0.03_{-0.01}^{+0.01}$ & $-1.60_{-0.18}^{+0.14}$ & $3.93_{-0.03}^{+0.03}$ & $3.69_{-0.01}^{+0.01}$ & $-0.65_{+0.03}^{-0.03}$ & $-4.65_{+0.30}^{-0.33}$   \\\\
SN2009jf$^{*}$ & $0.39_{-0.12}^{+0.19}$ & $-1.14_{+0.11}^{-0.20}$ & $0.13_{-0.04}^{+0.04}$ & $1.60_{-1.64}^{+1.52}$ & $3.95_{-0.02}^{+0.02}$ & $3.61_{-0.01}^{+0.01}$ & $-0.77_{+0.04}^{-0.03}$ & $-4.53_{+0.20}^{-0.23}$\\\\
\hline
iPTF13bvn & $0.15_{-0.09}^{+0.13}$ & $-1.12_{-0.13}^{+0.09}$ & $0.04_{-0.01}^{+0.01}$ & $-1.77_{-0.13}^{+0.11}$ & $3.83_{-0.02}^{+0.02}$ & $3.65_{-0.01}^{+0.01}$ & $-0.74_{+0.03}^{-0.03}$ & $-3.85_{+0.28}^{-0.30}$   \\\\
iPTF13bvn$^{*}$ & $-0.04_{-0.12}^{+0.21}$ & $-1.36_{+0.14}^{-0.18}$ & $0.12_{-0.04}^{+0.05}$ & $1.55_{-1.62}^{+1.54}$ & $3.82_{-0.01}^{+0.01}$ & $3.56_{-0.01}^{+0.00}$ & $-0.66_{+0.02}^{-0.02}$ & $-1.92_{+0.07}^{-0.08}$\\\\
\hline
SN2015ap & $0.19_{-0.03}^{+0.03}$ & $-0.95_{+0.03}^{-0.03}$ & $0.04_{-0.00}^{+0.01}$ & $-1.40_{-0.04}^{+0.04}$ & $4.06_{-0.02}^{+0.02}$ & $3.69_{-0.01}^{+0.01}$ & $-0.94_{+0.03}^{-0.03}$ & $-3.27_{+0.19}^{-0.20}$   \\\\
SN2015ap$^{*}$ & $0.25_{-0.12}^{+0.16}$ & $-0.24_{+0.15}^{-0.22}$ & $0.10_{-0.03}^{+0.05}$ & $1.73_{-1.58}^{+1.48}$ & $4.45_{-0.03}^{+0.03}$ & $3.41_{-0.25}^{+0.24}$ & $-0.80_{+0.07}^{-0.05}$ & $-3.64_{+0.29}^{-0.40}$\\\\
\hline
SN2016bau & $0.19_{+0.02}^{-0.01}$ & $-0.94_{+0.02}^{-0.02}$ & $0.05_{-0.00}^{+0.00}$ & $-1.46_{-0.06}^{+0.06}$ & $3.95_{-0.02}^{+0.02}$ & $3.82_{-0.01}^{+0.01}$ & $-0.65_{+0.02}^{-0.02}$ & $-4.88_{+0.14}^{-0.06}$   \\\\
SN2016bau$^{*}$ & $-0.51_{+0.23}^{-0.18}$ & $-1.01_{+0.21}^{-0.25}$ & $0.10_{-0.03}^{+0.04}$ & $2.45_{-3.41}^{-2.48}$ & $3.58_{-0.02}^{+0.14}$ & $3.55_{-0.38}^{+0.31}$ & $-1.72_{+0.23}^{-0.18}$ & $-4.41_{+0.29}^{-0.36}$\\\\
\hline

\hline
\end{tabular}
\end{scriptsize}
\caption{Best-fit parameters and 68\,\% uncertainties for the {\tt default} model. The parameters for the source with * are the results from \citet[][]{Meyer2020} presented here for comparison with our studies. In this table, $M_{ej}$ is the ejecta mass, $f_{Ni}$ is the Nickel mass fraction, $\kappa$ is the Thomson electron scattering opacity, $\kappa_{\gamma}$ is gamma-ray opacity of the SN ejecta, $v_{ej}$ represents the ejecta velocity, $T_{min}$ is an additional parameter for temperature floor [ please see \citet[][]{Nicholl2017}, for further details ], $\sigma$ is an additional variance parameter which is added to each uncertainty of the measured magnitude so that the reduced $\chi^{2}$ approaches 1, and $t_{exp}$ is the epoch of explosion since first detection.  
\label{tab:mosfit_default}}
\end{table*}




\section{Understanding the possible progenitors of the sub-sample of CCSNe using {\tt MESA}} 
\label{Mesa}
{\tt MESA} (version r11701) is a one dimensional stellar evolution code. It is open source, rich, efficient having thread-safe libraries for a wide range of applications in computational stellar astrophysics. The capacity of {\tt MESA} \textbf{is} enormous. It can be used to study various phases of stellar evolution resulting in various types of SNe, pulsations in stars, accretion onto a Neutron star, black hole formations and many other astrophysical phenomena. Following \citet[][]{Cao2013}, \citet[][]{Aryan2021a}, and \citet[][]{Sahu2011}, iPTF13bvn, SN~2015ap, SN~2016bau, and SN~2009jf have progenitors with ZAMS masses in the range of 11 to 20\,M$_{\odot}$. In this work, following \citet[][]{Aryan2021a} and \citet[][]{Pandey2021}, we attempt to understand the physical and chemical properties of 12\,M$_{\odot}$ and 20\,M$_{\odot}$ ZAMS stars which could be the possible ZAMS mass range for the progenitors of these SNe. We briefly mention the {\tt MESA} settings and assumptions for our calculations below.

\begin{figure}
\centering
\includegraphics[width=\columnwidth]{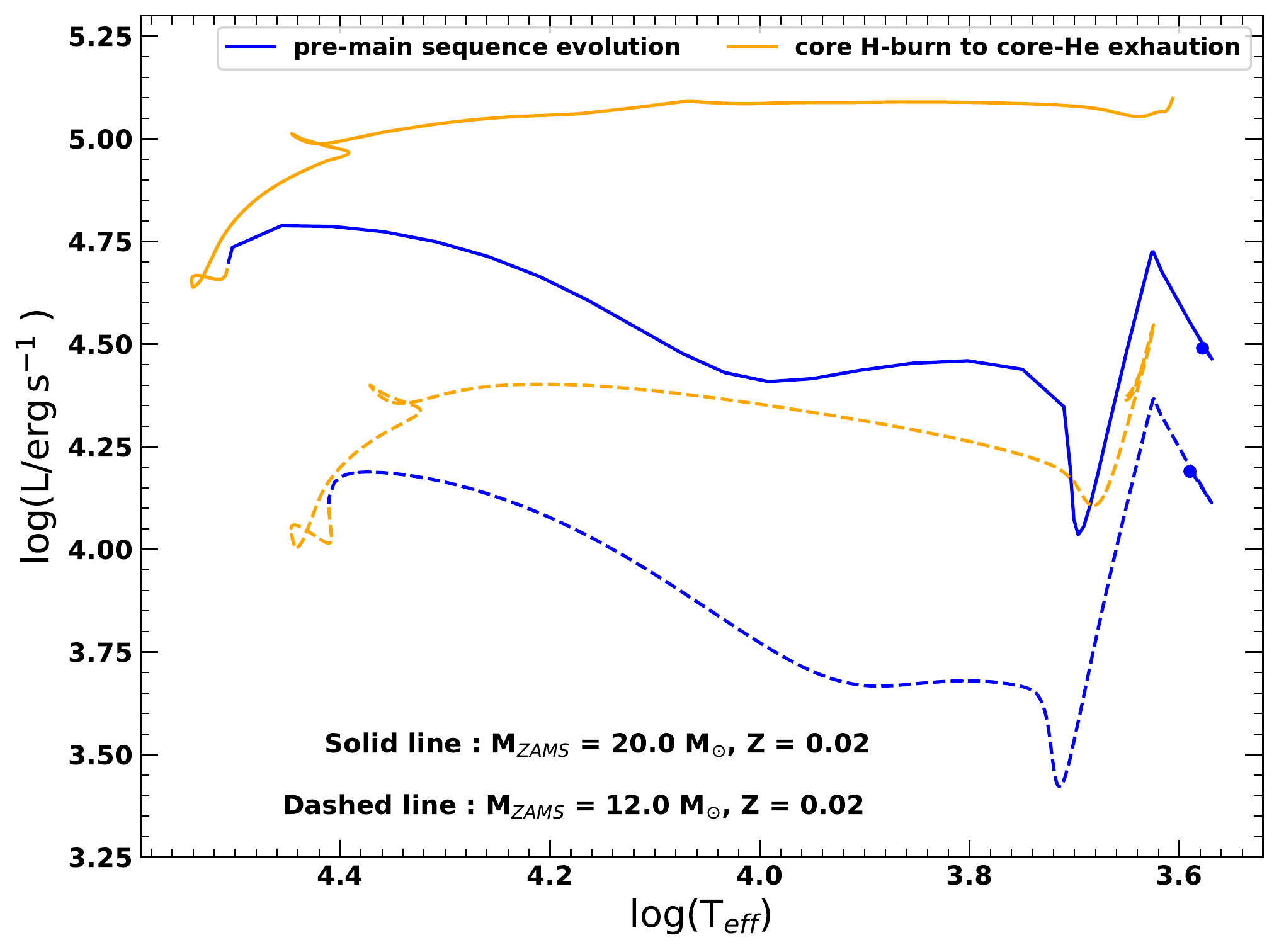}
\caption{The evolutions of 12\,M$_{\odot}$ and 20\,M$_{\odot}$ models (both having Z = 0.02) on HR diagram from PMS till the exhaustion of He-burning in their core. The blue solid circles mark the beginning of PMS evolution of the two models. }
\label{fig:HR_diagram}
\end{figure}

\begin{figure*}
\centering
    \includegraphics[height=8.0cm,width=8.5cm,angle=0]{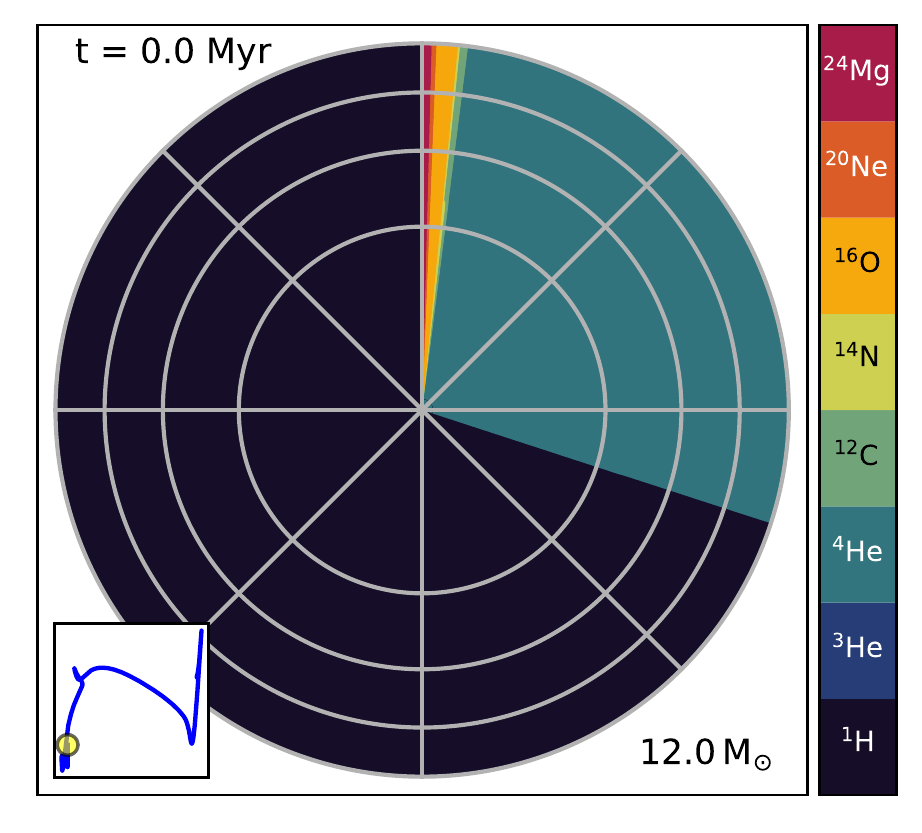}
    \includegraphics[height=8.0cm,width=8.5cm,angle=0]{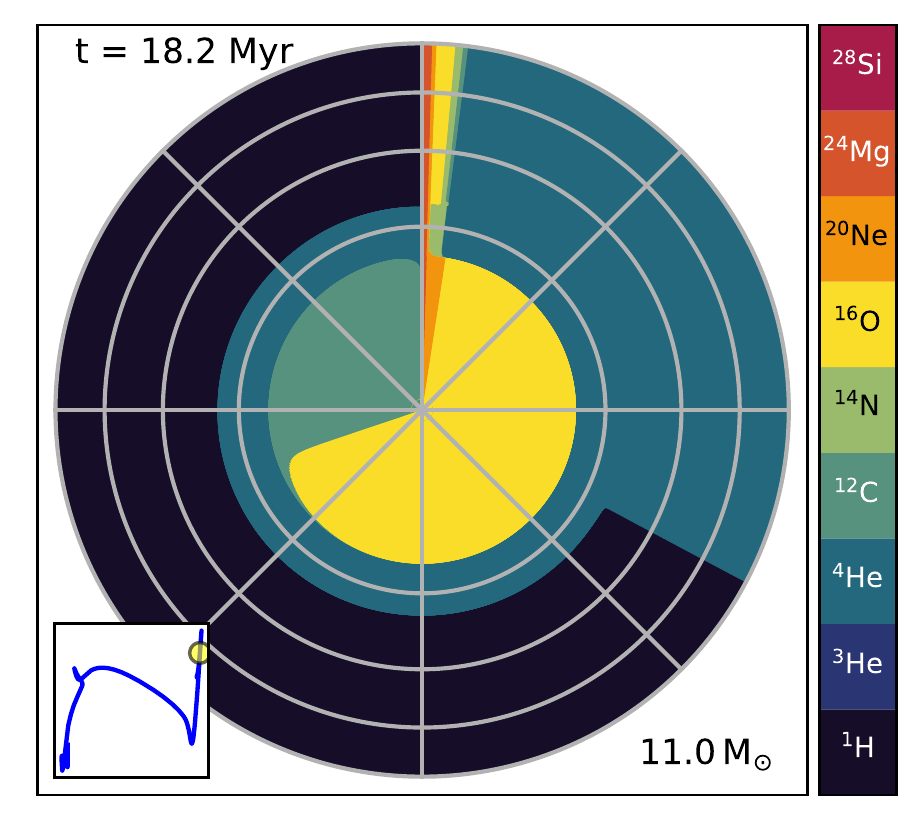}
   \caption{The abundances of various elements in the stellar interior of the 12\,M$_{\odot}$ progenitor with Z = 0.02, at two stages. {\em Left:} The abundances of various elements when the model has just arrived on the main-sequence. {\em Right:} The abundances of various elements as the model has finished core-He burning. The compositions of heavier elements in the stellar interior have increased now.}
    \label{fig:abundance_12msun}
\end{figure*}

\begin{figure*}
\centering
    \includegraphics[height=8.0cm,width=8.5cm,angle=0]{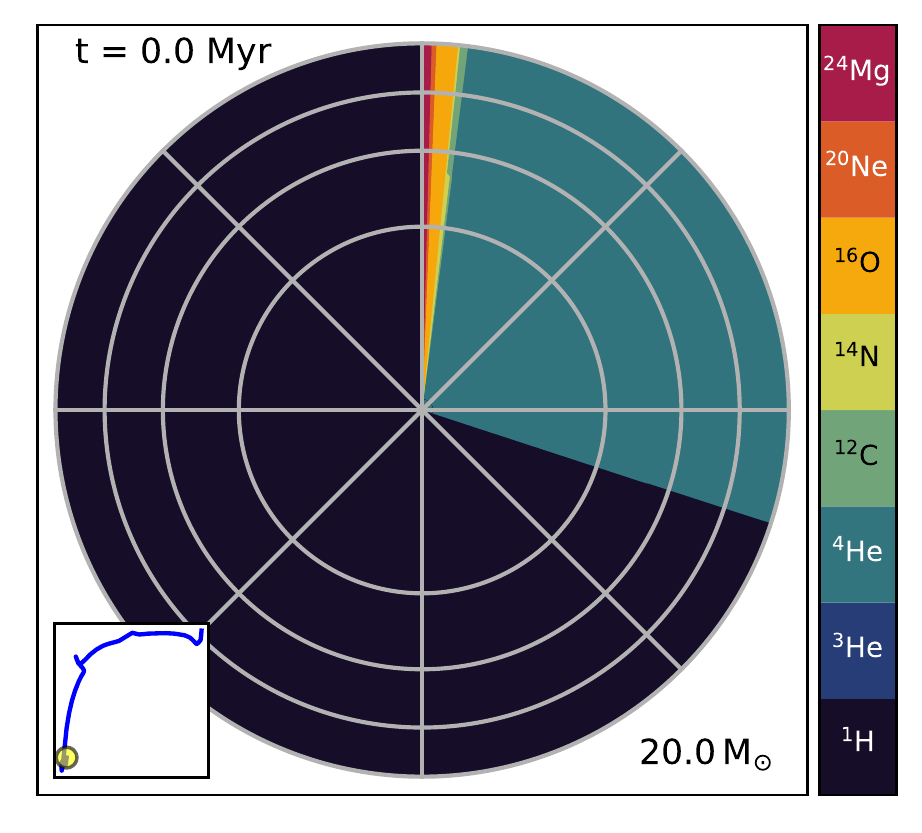}
    \includegraphics[height=8.0cm,width=8.5cm,angle=0]{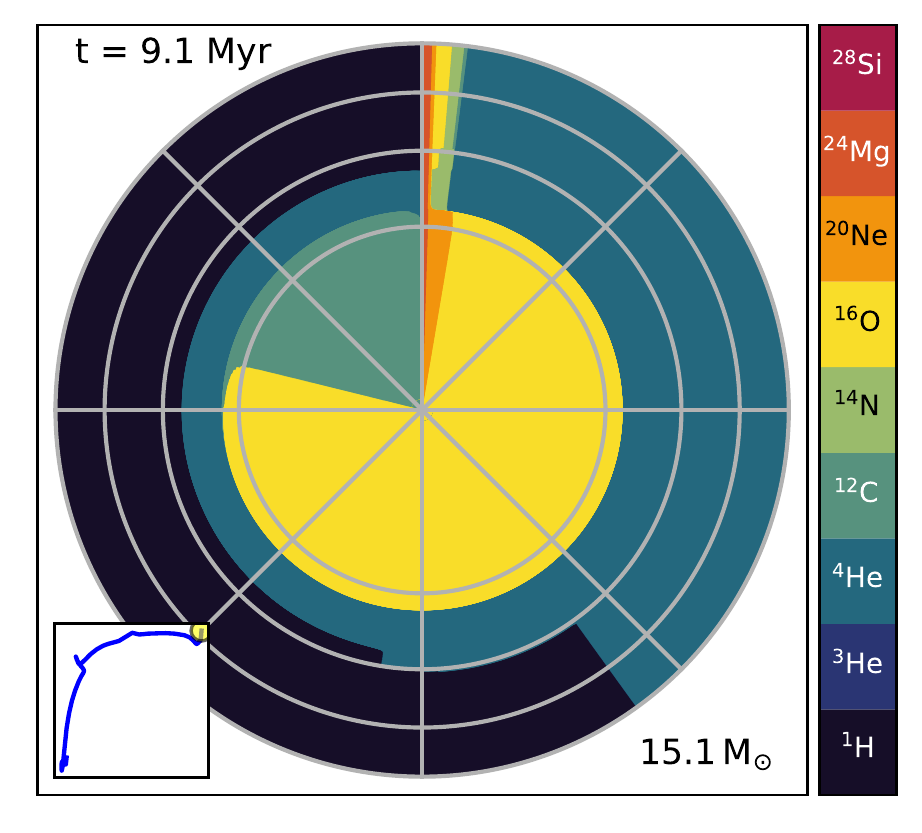}
   \caption{The abundances of various elements in the stellar interior of the 20\,M$_{\odot}$ progenitor with Z = 0.02, at two stages. {\em Left:} The abundances of various elements when the model has just arrived on the main-sequence. {\em Right:} The abundances of various elements as the model has finished core-He burning. Similar to 12\,M$_{\odot}$ model, the compositions of heavier elements have increased now.}
    \label{fig:abundance_20msun}
\end{figure*}

For the 12\,M$_{\odot}$ ZAMS progenitor model that could be the progenitor of iPTF13bvn, SN~2015ap or SN~2016bau (from \citet[][]{Cao2013}, \citet[][]{Aryan2021a}), our calculations closely follow \citet[][]{Aryan2021a}.  A brief description of the 12 M$_{\odot}$ ZAMS progenitor model has been provided. Starting from the pre--main sequence (PMS), the 12\,$M_{\odot}$ ZAMS star is evolved through various stages on the HR diagram until the onset of core-collapse. We consider rotation--less progenitor having an initial metallicity of $Z = 0.02$  because these three SNe prove to be arising in the regions having metallicities close to solar metallicity. The convection is modelled using the mixing theory of \citet[][]{Henyey1965} by adopting the Ledoux criterion. The mixing-length parameter is set to $\alpha = 3.0$ in the region where the mass fraction of hydrogen is greater than 0.5, and $\alpha = 1.5$ for the other regions. Further, the semi-convection is modelled following \citet[][]{Langer1985} having an efficiency parameter of $\alpha_{\mathrm{sc}} = 0.01$. For the thermohaline mixing, following \citet[][]{Kippenhahn1980},  the efficiency parameter is set  as $\alpha_{\mathrm{th}} = 2.0$. The convective overshooting is modelled using the  diffusive approach of \citet[][]{Herwig2000}, with $f= 0.01$ and $f_0 = 0.004$ for all the convective core and shells. For the stellar wind, {\tt Dutch} scheme is used with a scaling factor of 1.0. We employed satisfactory spatial and temporal resolution in our models by choosing {\tt mesh\_delta\_coeff} = 1.0 and {\tt varcontrol\_target} = 5d-4.

For the 20\,M$_{\odot}$ ZAMS progenitor model, similar settings have been used including an initial metallicity of $Z = 0.02$. SN~2009jf occurred in a region having metallicity close to the solar metallicity thus $Z = 0.02$ is used for the 20\,M$_{\odot}$ model too, serving as the possible progenitor of SN~2009jf. A slightly better spatial resolution is  employed by taking {\tt mesh\_delta\_coeff} = 0.8.

Figure~\ref{fig:HR_diagram} shows the evolutions of the 12\,M$_{\odot}$ and 20\,M$_{\odot}$ ZAMS progenitors on the HR diagram starting from pre-main-sequence (PMS) to the exhaustion of He-burning in the core. At the end of the core He-burning phase, both models are living in the red-giant/supergiant phase. Figure~\ref{fig:abundance_12msun} and figure~\ref{fig:abundance_20msun} show the snapshots of the chemical compositions of the progenitor stars at two phases each. The left panels of figure~\ref{fig:abundance_12msun} and figure~\ref{fig:abundance_20msun} show the stellar compositions when the models have just landed on the ZAMS while their right panels show the chemical compositions of models until the end of the core He-burning. The grey circles in each subplot indicate the location of the mass coordinates at 0.25, 0.5. 0.75 and 1.00 times the total stellar mass. It can be noticed that initially ( depending on the metallicity ), the fraction of H and He are much higher than other heavy metals but as the models evolve and reach the end of the He-burning in the core, the composition of heavier elements increases in the core. Other important noticeable properties include the significant stripping of envelope due to the presence of stellar wind as the models evolve. The 12\,M$_{\odot}$ progenitor model has lost around 1\,M$_{\odot}$ as it reaches the termination stage of core-He burning while at the similar stage, the 20\,M$_{\odot}$ progenitor model has lost a significant amount of envelope retaining 15.1\,M$_{\odot}$ of total ZAMS mass. The high mass loss in the case of 20\,M$_{\odot}$ progenitor model could be attributed to the presence of comparatively stronger stellar winds than the case of 12\,M$_{\odot}$ model.

\begin{figure}
\centering
\includegraphics[width=\columnwidth]{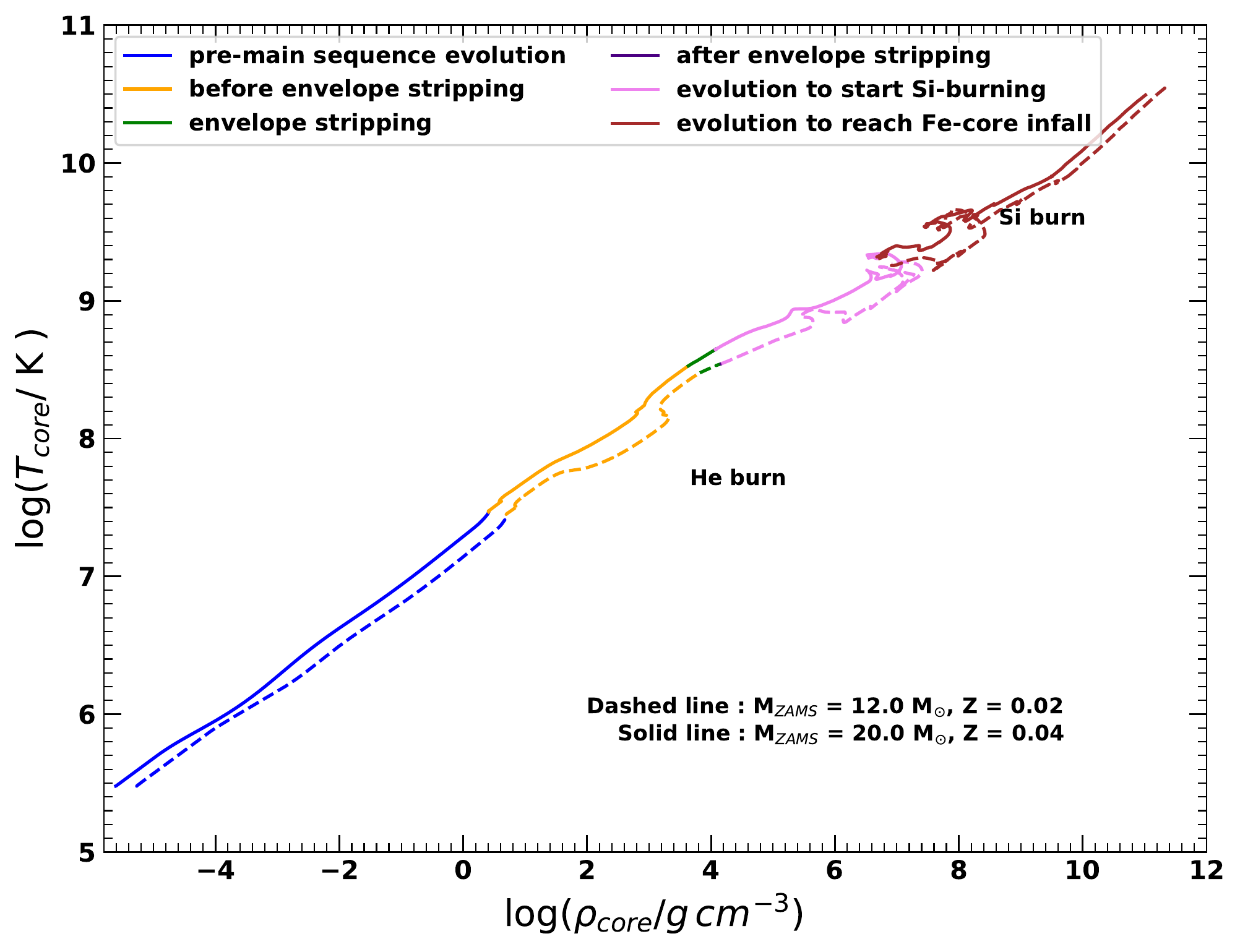}
\caption{The variation of core temperature with core density as the models evolve through various phases on the HR diagram. Notice the very high core temperature and core density of the order of 10$^{10}$\,K and 10$^{10}$\,gm\,cm$^{-3}$, respectively, towards the last evolutionary phases. Such high core temperatures and densities are indicative of the arrival of the core-collapse phase.  }
\label{fig:Rhoc_vs_Tc}
\end{figure}

\begin{figure*}
\centering
    \includegraphics[height=8.0cm,width=8.5cm,angle=0]{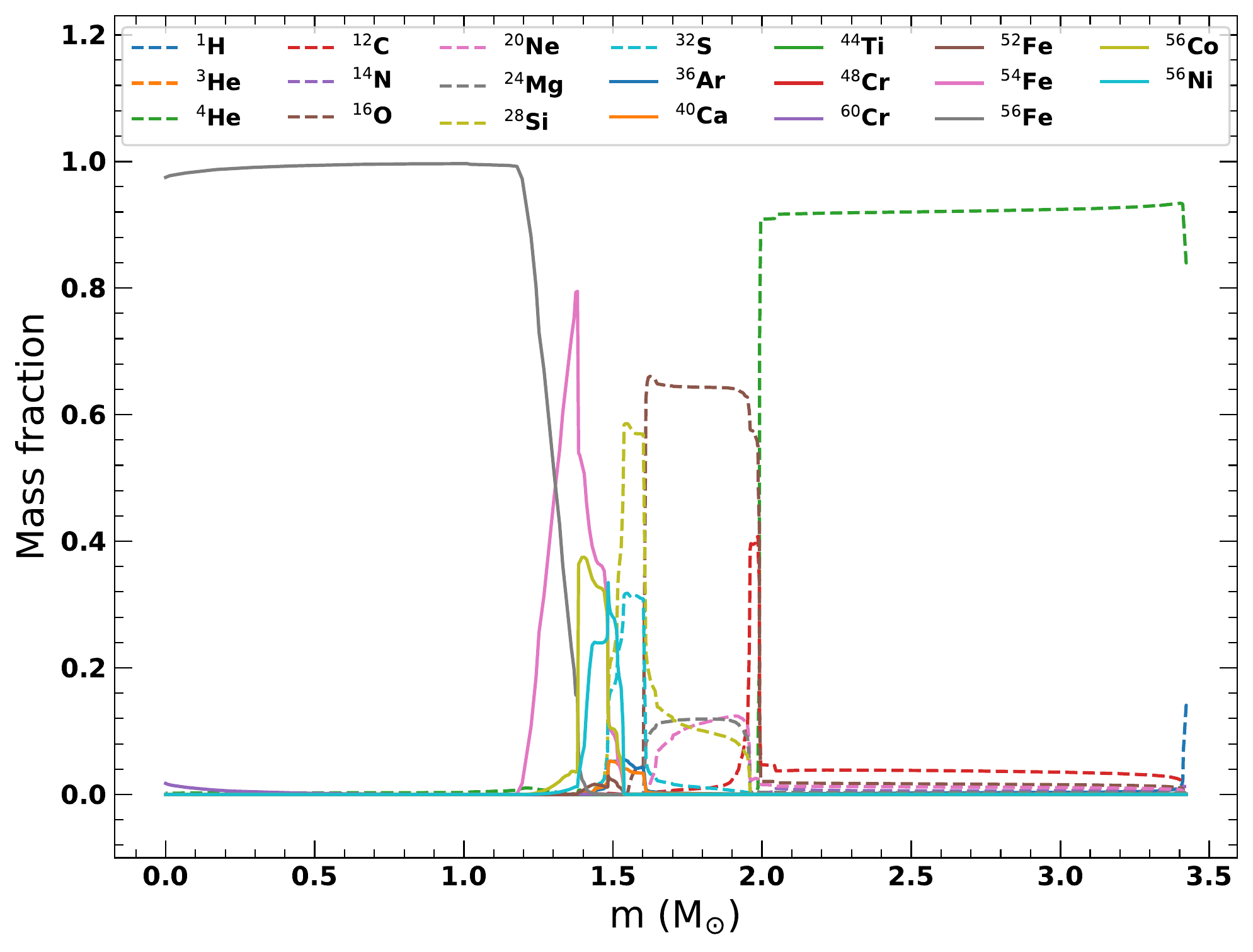}
    \includegraphics[height=8.0cm,width=8.5cm,angle=0]{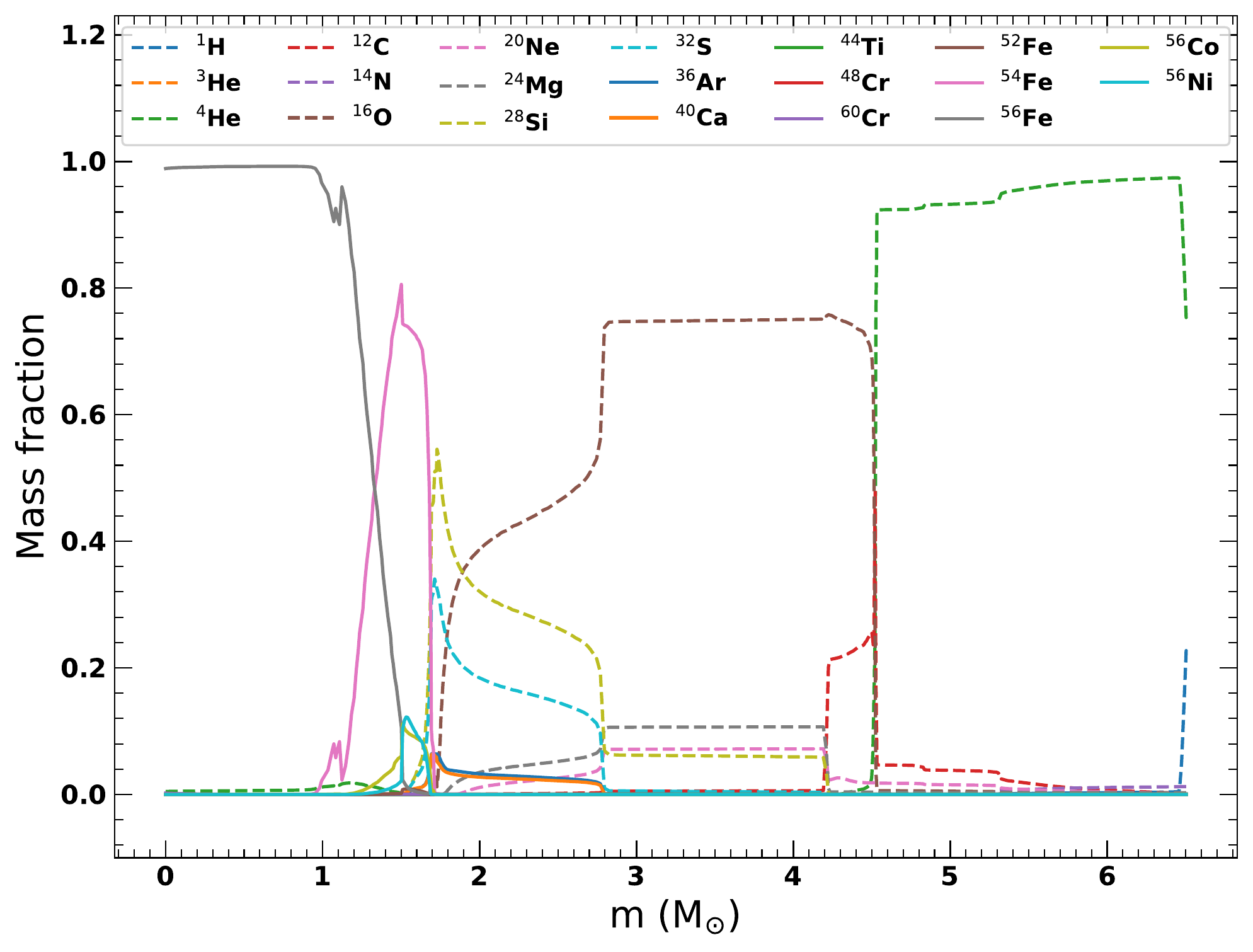}
   \caption{ The mass fraction of various elements in the stellar interiors near the onset of core-collapse. Both the models show very high mass fraction of $^{56}$Fe in the core, which is indicative of the arrival of the core-collapse phase. {\em Left:} The mass fraction of various elements near the arrival of the core-collapse phase of a 12\,M$_{\odot}$ ZAMS progenitor with metallicity Z=0.02. {\em Right:} The mass fraction of various elements near the arrival of the core-collapse phase of a 20\,M$_{\odot}$ ZAMS progenitor with metallicity Z = 0.02. }
    \label{fig:mass_fraction}
\end{figure*}

The SNe considered in this study are all type Ib. SNe~Ib have been considered to originate from massive stars which lose almost all of their hydrogen envelope, most probably due to binary interaction \citep[e.g.,][]{Yoon2010, Dessart2012, Eldridge2016} or due to strong stellar winds \citep[e.g.,][]{Gaskell1986, Eldridge2011, Groh2013}. Here, to produce such a stripped model, the hydrogen envelopes are artificially stripped. Specifically, after evolving the model until the exhaustion of helium, an artificial mass-loss rate of $\dot{M} \gtrsim 10^{-4}$\,M$_{\odot}\,\mathrm{yr}^{-1}$ has been imposed until the total hydrogen masses of the models go down to 0.01\,M$_{\odot}$. After the hydrogen masses in the models reach the specified limit, the artificial mass loss is switched off and the models are further evolved on the HR diagram until the onset of the core-collapse. Starting from 12 M$_{\odot}$ at ZAMS, our model has a total mass of 3.42\,M$_{\odot}$ at an stage just before the core--collapse. For the 20 M$_{\odot}$ ZAMS progenitor, the model has a total mass of around 6.50\,M$_{\odot}$ just before the core-collapse.

Thus, after approaching the ZAMS sequence, our models evolve to become giants/supergiants. Further, they suffer stripping and evolve ahead to start Si-burning in their respective cores. As a result, our models develop inert Fe-cores that result in the core-collapse due to the absence of any further fusion processes in the core. Figure~\ref{fig:Rhoc_vs_Tc} shows the variation of core temperature ($T_{core}$) with the core density ($\rho_{core}$) as the models evolve from PMS until the onset of their core-collapses. In the last evolutionary phases, the $T_{core}$ and $\rho_{core}$ reach in excess of  10$^{10}$\,K and 10$^{10}$\,gm\,cm$^{-3}$, respectively. Such high central temperatures and densities are considered to be the suitable physical conditions for the stellar core to collapse.

Further, figure~\ref{fig:mass_fraction} shows the mass fractions of various elements present inside the model stars when their cores are about to collapse. Near the surface of the stellar models, the fraction of He is much higher compared to other elements. Such high mass fractions of He near the surface of progenitors just before the core collapse is responsible for the type Ib SNe displaying strong He-features in their spectra. As we move inwards towards the centre of the stellar models, the cores consist mainly inert $^{56}$Fe, responsible for the cores to collapse.

\section{Results and Discussion}
\label{Results}
This work demonstrated the usefulness of publicly available analysis tools to understand the physical and chemical properties of a sub-set of CCSNe and their possible progenitors. We used publicly available data as inputs to {\tt MOSFiT}. Utilizing these data, {\tt MOSFiT} provided various physical and chemical properties of CCSNe for assumed progenitor stars along with explosion epochs and range of required temperatures, velocities, ejecta mass, opacity, etc. Further, based on these observed properties, a certain ZAMS mass progenitor could be chosen as the possible progenitors of these CCSNe and the stellar evolutions depicting various stages of evolutions could be performed to shed light on the physical structure and required chemical engineering. The findings of the present studies can be summarized as mentioned below: 

1) With the help of {\tt MOSFiT}, we fit the multi-band light curves of four H-stripped CCSNe namely, SN2009jf, iPTF13bvn, SN2015ap, and SN2016bau. The parameters obtained through {\tt MOSFiT} fittings were compared to those available in the literature. We demonstrated the usefulness of {\tt MOSFiT} and how it could be used with an extensive data set to constrain various physical parameters more realistically. 

2) In the later part of this study, we demonstrated the importance of {\tt MESA} to understand the physical and chemical properties of possible progenitors. {\tt MESA} proved to be an excellent tool to study stellar evolution. In this work, we performed 1-dimensional stellar evolutions of two progenitor models having ZAMS masses of 12\,M$_{\odot}$ and 20\,$_{\odot}$, which could serve as the possible progenitors of the four H-stripped CCSNe considered for the present study. We studied the evolutions of these models on HR diagram as they evolved through various phases throughout their lifetime. Further, we studied the variation of the chemical composition inside the stellar interior as the models evolved on the main--sequence and reached the stage of He-burning termination in the core. It was noticed that as the model evolved on the main--sequence and reached the stage of termination of He-burning in the core, the stellar interior composed more and more of heavier metals.

3) Further, we studied the variation of $\rho_{core}$ and $T_{core}$ as the models evolved from PMS upto the stages where their cores undergo core--collapse. The $\rho_{core}$ and $T_{core}$ reached in excess of 10$^{10}$\,gm\,cm$^{-3}$ and 10$^{10}$\,K in the late evolutionary stages marking the onset of core--collapse.

4) As another piece of evidence, we also studied the mass fractions of various elements in the stellar interiors. We found out that during the last evolutionary stages, the central regions of the stellar models are mainly composed of inert $^{56}$Fe, marking the arrival of core--collapse stage. 

5) As a next step, the output of {\tt MESA} models on the verge of the onset of core--collapse could be provided as input to other explosion codes capable of simulating synthetic stellar explosions. The outputs obtained through such simulations could be compared to actual SNe properties and to understand new types of transients in near future. 

\section{Conclusion}
\label{Conclusion}
In this work, we demonstrated the significance of {\tt MOSFiT} and {\tt MESA} to understand the physical and chemical properties of H-stripped CCSNe, particularly Type Ib. {\tt MOSFiT} is used the fit the multi-band light curves of SNe by taking into account {\tt default} powering mechanisms. The fitting results provide various physical properties including the SN temperature, velocity, opacity, ejecta mass, explosion epochs, etc. Depending on the high or low ejecta mass, ZAMS progenitors of different initial masses could be modeled. Also, the variations in opacity, explosion epochs and photospheric velocities are highly sensitive to  SN light curves. Thus, parameters obtained using {\tt MOSFiT} could serve as initial guesses for the progenitor models using {\tt MESA} that later explode synthetically to give SNe light curves and photospheric velocities. So, based on these properties, stars of certain ZAMS mass range can be modeled to depict as possible progenitors of CCSNe using {\tt MESA} and can be evolved from PMS up to the onset of core--collapse using {\tt MESA}. The snapshots of various physical and chemical properties can be obtained from {\tt MESA} outputs which are extremely essential to understand the stellar properties of possible progenitors of CCSNe. Thus, our studies display how publicly available analysis tools can be used to remove the shear dependency on unpublished data to extract useful scientific information about a variety of transients and to understand related aspects of nuclear astrophysics, a broader and interdisciplinary emerging research area.

\section*{Acknowledgements}
We are thankful to the referee for providing valuable comments that were highly helpful in improving the manuscript further. A.A. acknowledges funds and assistance provided by the Council of Scientific \& Industrial Research (CSIR), India with file no. 09/948(0003)/2020-EMR-I. RG and SBP acknowledge the financial support of ISRO under AstroSat archival Data utilization program (DS$\_$2B-13013(2)/1/2021-Sec.2). We further acknowledge that the High Performance Computing (HPC) facility at ARIES played extensive role in carrying out the simulations performed in this work. 

\section*{Additional Software}
NumPy \citep[][]{Harris2020}, Matplotlib \citep[][]{Caswell2021}, mesaPlot \citep[][]{Wise2019}, Mesa\_Reader \citep[][]{Bill2017}, TULIP \citep[][]{Laplace2021}
%





\end{document}